\documentclass[12pt,a4paper,dvips]{article}

\usepackage{a4p}
\usepackage{graphicx}

\newlength{\figwidth}
\begin{document}
\setlength{\unitlength}{1mm}
\thispagestyle{empty}
\begin{center}
{\Large EUROPEAN ORGANIZATION FOR NUCLEAR RESEARCH}
\end{center}
\begin{flushright}
  {\large CERN-PH-EP/2004-030 }\\
  {\large July 6, 2004}\\
\end{flushright}  

\vspace{2cm}
\begin{center}
{\Large\bf Physics of W bosons at LEP}
\vspace{2cm}

{\bf\large  Salvatore Mele}\\
{ CERN, CH-1211, Geneva 23, Switzerland\\
 INFN, Sezione di Napoli, I-80126, Italy.\\
\it{Salvatore.Mele@cern.ch}}

\end{center}
\vspace{2cm}

\begin{abstract}

The high-energy and high-luminosity data-taking campaigns of the LEP
$\rm e^+e^-$ collider provided the four collaborations, ALEPH, DELPHI,
L3 and OPAL, with about 50\,000 W-boson pairs and about a thousand
singly-produced W bosons. This unique data sample has an unprecedented
reach in probing some aspects of the Standard Model of the electroweak
interactions, and this article reviews several achievements in the
understanding of W-boson physics at LEP.  The measurements of the
cross sections for W-boson production are discussed, together with
their implication on the existence of the coupling between Z and W
bosons. The precision measurements of the magnitude of triple
gauge-boson couplings are presented. The observation of the
longitudinal helicity component of the W-boson spin, related to the
mechanism of electroweak symmetry breaking, is described together with
the techniques used to probe the CP and CPT symmetries in the W-boson
system.  A discussion on the intricacies of the measurement of the
mass of the W boson, whose knowledge is indispensable to test the
internal consistency of the Standard Model and estimate the mass of
the Higgs boson, concludes this review.

\end{abstract}

\vspace{1.5cm}

\begin{center}
  {\it To appear in a special issue of Physics Reports celebrating
the 50$^{th}$ anniversary of CERN}
\end{center}
\newpage

%%%%%%%%%%%%%%%%%%%%%%%%%%%%%%%%%%%%%%%%%%%%%%%%%%%%%%%%%%%%%%%%%%
\section*{Introduction}
%%%%%%%%%%%%%%%%%%%%%%%%%%%%%%%%%%%%%%%%%%%%%%%%%%%%%%%%%%%%%%%%%%

The mission of the LEP project was to further the understanding of the
Standard Model of the electroweak interactions, and the study of W
bosons is a unique tool to meet this challenge. The study of W-boson
pair-production probes two cornerstones of the Standard Model, namely
the existence of the coupling between Z and W bosons and of the
longitudinal helicity component of the W-boson spin. The mass of the W
boson, $m_{\rm W}$, is a free parameter of the Standard Model and its
measurement is therefore indispensable. Precise knowledge of $m_{\rm
W}$ allows, through the mechanism of radiative corrections, to test
the internal consistency of the Standard Model, as described elsewhere
in this Report. In addition, it provides hints on the mass of the
yet-unobserved Higgs boson.

As from 1996, the LEP centre-of-mass energy, $\sqrt{s}$, was
steadily increased above the W-boson pair-production threshold of $2
m_{\rm W} \approx 161$ GeV, opening a window on W-boson physics at
LEP. By the machine shut-down in the year 2000, a total of about
2.7~fb$^{-1}$ of integrated luminosity had been delivered to the four
detectors, and around 50\,000 events with W-boson pairs were on
tape. 

\begin{figure}
  \begin{center}
    \includegraphics[width=\textwidth]{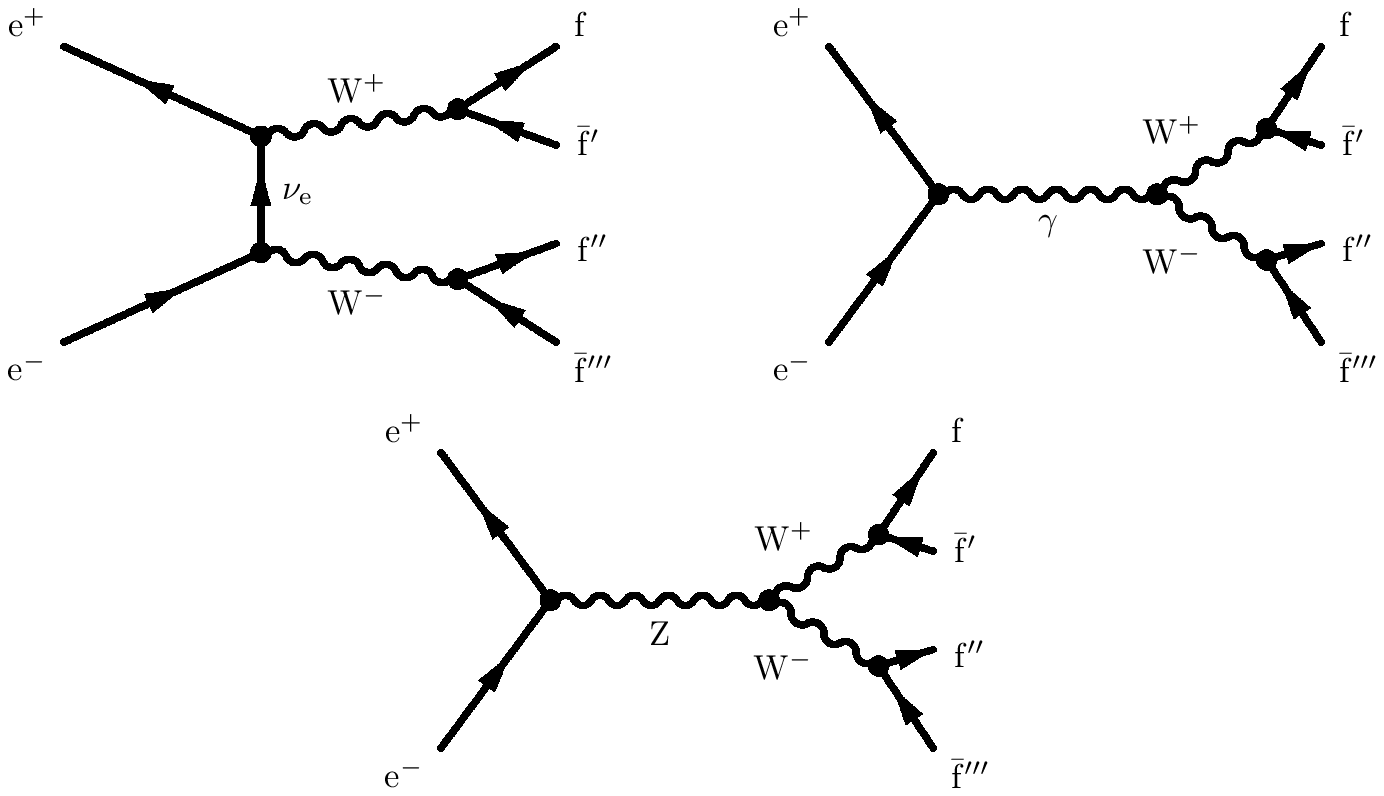}
    \caption{\label{fig:W1} Feynman diagrams describing W-boson pair-production
     at LEP.}
  \end{center}
\end{figure}

W-boson pair-production proceeds through $t$-channel neutrino exchange
and $s$-channel annihilation with the mediation of either a photon or
a Z boson, as shown in Figure~\ref{fig:W1}. The $s$-channel diagrams
are sensitive to the triple gauge-boson couplings (TGCs) $\gamma$WW
and ZWW~\cite{yellow,Grunewald:2000ju}.  More than a thousand W bosons
were also single-produced at
LEP~\cite{Grunewald:2000ju,yellow4f}. This is a particular case of the
generic four-fermion process $\rm e^+e^-\rightarrow e^+ \nu_e f
\bar{f}'$\footnote{Unless specified otherwise, charge conjugate
processes are implied throughout this Article.}, described by diagrams
such as those depicted in Figure~\ref{fig:W2}.

The identification of events from W-boson pair-production is discussed
in the following sections, together with the determination of the
cross section of this process and of W-boson branching fractions.
Results from the study of single W-boson production are also
presented. Details are then given on the investigation of W-boson
polarisation and TGCs. Finally, the measurement of $m_{\rm W}$ is
described. Additional details on those topics and a comprehensive list
of references can be found in Reference~\cite{unknown:2003ih}. At the
time of writing, a few of these results are still
preliminary. Nonetheless, as complex data analyses are nearing
conclusion, final results are expected not to show large differences
with those discussed below.

%%%%%%%%%%%%%%%%%%%%%%%%%%%%%%%%%%%%%%%%%%%%%%%%%%%%%%%%%%%%%%%%%%
\section*{W-boson pair-production}
%%%%%%%%%%%%%%%%%%%%%%%%%%%%%%%%%%%%%%%%%%%%%%%%%%%%%%%%%%%%%%%%%%

W bosons decay either into hadrons or into a charged lepton and a
neutrino, with branching fractions of 67.5\% and 32.5\%,
respectively. Therefore, W-boson pair-production is observed in three
different topologies:
\begin{itemize}
\item[-] ``Fully-hadronic events'', for 45.5\% of the pairs, in which
  both W bosons decay into hadrons with a signature of four hadronic
  jets in the detectors;
\item[-] ``Semi-leptonic events'', for 43.9\% of the pairs, in which
  only one W boson decays into hadrons, resulting in events with two
  hadronic jets, a lepton and missing energy and momentum due to an
  undetected neutrino;
\item[-] ``Fully-leptonic events'', for 10.6\% of the pairs, in which
  both W bosons decay into leptons with a signature of just two
  charged leptons and large missing energy and momentum carried away
  by the two neutrinos.
\end{itemize}
The LEP collaborations developed techniques to select these events
with the highest possible efficiency, while suppressing the competing
background from other Standard Model processes~\cite{sigmaW}.

Fully-hadronic events are selected with multivariate analyses which
rely on event-shape information discriminating four-jet events from
two-jet events and on variables quantifying the compatibility of the
event kinematics with the production of two W bosons. Three
pairings of four jets into two W bosons are possible and the one
which best fits the W-boson pair-production
hypothesis is retained. Efficiencies of about 80\% are reached, for a residual
background of about 15\%, mostly due to four-jet events originating
from higher-order contributions to the process $\rm e^+e^- \rightarrow
q\bar{q}$.

Semi-leptonic events are selected by requiring the invariant masses of
the two-jet and lepton-neutrino systems to be compatible with $m_{\rm
W}$.  The neutrino four-momentum is deduced from the measured jet and
lepton momenta by imposing energy-momentum conservation. Selection
criteria on the energy of the charged lepton, the transverse momentum
of the event and the direction of the missing momentum reduce
the background from $\rm e^+e^- \rightarrow q\bar{q}$ events
containing leptons and from four-fermion processes. Efficiencies
between 50\% and 90\% are achieved, the lower values corresponding to
tau leptons, which are more complex to reconstruct due to their
hadronic decays and the presence of additional neutrinos. The
background contamination varies from less than 5\% for muons up to
20\% for tau leptons and is due to two- and four-fermion processes.

Fully-leptonic events are tagged by the presence of two high-energy
charged leptons and large missing energy and momentum. Background from
the $\rm e^+e^- \rightarrow \ell^+\ell^-\gamma$ process, where the
initial-state-radiation photon escapes undetected along the beam pipe,
is reduced by requiring events with large transverse momentum and a
missing momentum pointing away from the beam axis. Selection
efficiencies vary from 30\% to 70\%, depending on the lepton flavour,
the lowest values corresponding to tau leptons. The background
contamination varies between 15\% and 30\% and is due to two- and
four-fermion processes.

\begin{figure}
  \begin{center}
    \includegraphics[width=\textwidth]{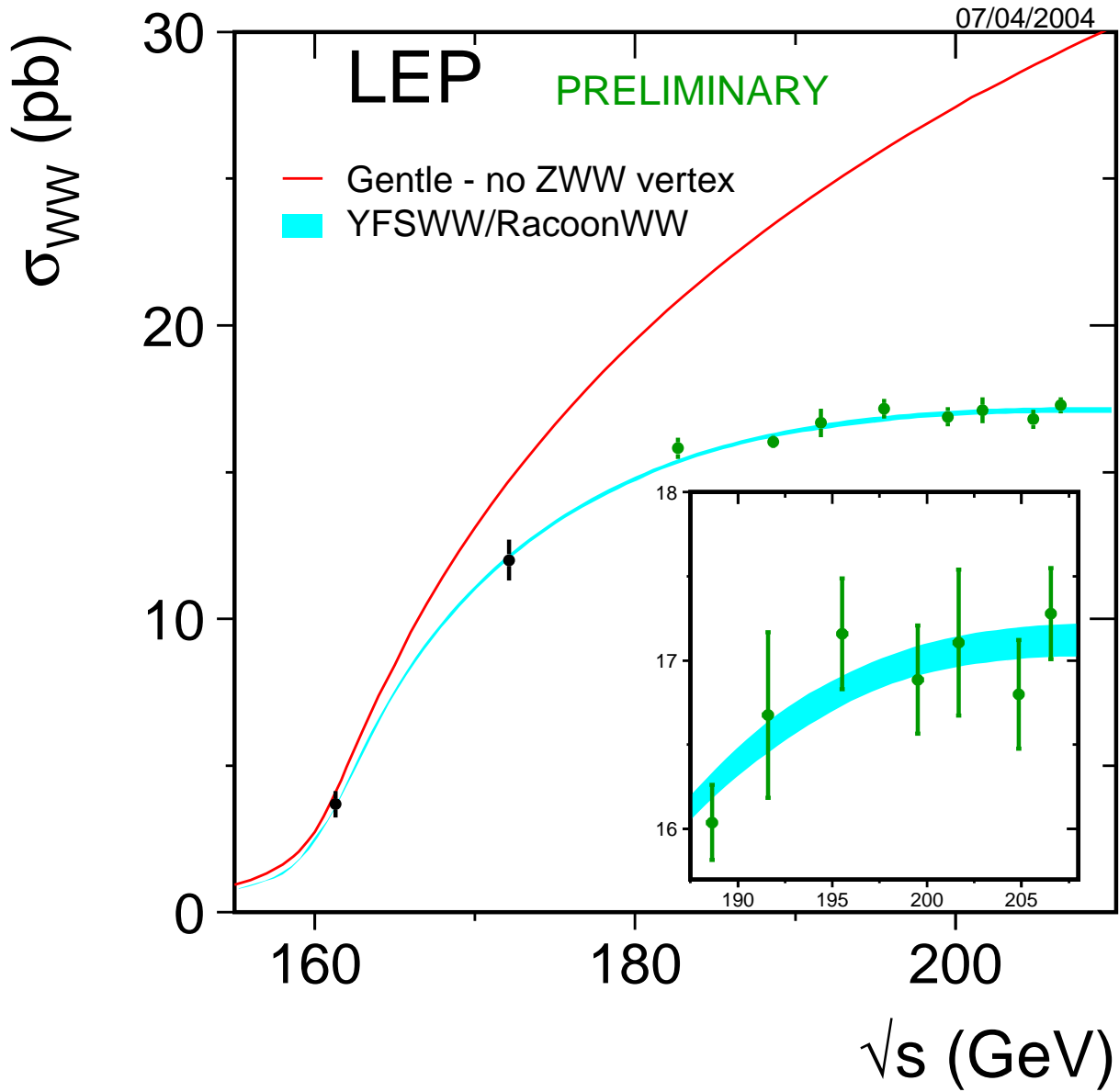}
    \caption{\label{fig:W3} Measurement of the cross section for
    W-boson pair-production at LEP as a function of $\sqrt{s}$. Values
    above 180~GeV are still preliminary. Standard Model
    predictions~\cite{sigmaWtheo} are indicated by the band, whose
    width represents the theoretical uncertainty of
    0.5\%~\cite{Grunewald:2000ju,sigmaWtheo}. Predictions in the
    absence of the ZWW couplings~\cite{Bardin:1996zz} are also shown.}
  \end{center}
\end{figure}

In total, around 40\,000 W-boson pairs are selected by the four
collaborations and the cross sections for W-boson pair-production are
measured~\cite{sigmaW}. The combined results~\cite{unknown:2003ih} are
presented in Figure~\ref{fig:W3} as a function of $\sqrt{s}$.  These
results establish the existence of the ZWW coupling, as a much higher
cross section would characterise its absence~\cite{Bardin:1996zz}.
The measurements are in excellent agreement with the Standard Model
predictions~\cite{sigmaWtheo} as quantified by the ratio, ${\cal
R}^{\rm WW}$, of the measured, $\sigma_{\rm WW}^{\rm meas}$, and the
expected, $\sigma_{\rm WW}^{\rm theo}$, cross sections:
\begin{displaymath}
{\cal R}^{\rm WW}={\sigma_{\rm WW}^{\rm meas} \over \sigma_{\rm
WW}^{\rm theo}}=0.997\pm 0.010.
\end{displaymath}
The uncertainty on ${\cal R}^{\rm WW}$ receives equal contributions
from statistical and systematic uncertainties. The latter are mainly
due to the description of QCD processes in both the signal and
background modelling.

Low values of ${\cal R}^{WW}$ were initially observed, calling for the
present improved description of this
process~\cite{Grunewald:2000ju}. The so-called leading-pole and
double-pole approximations were developed to take into account the
exchange of a virtual photon between the particles involved in the
process.  In addition, an improved treatment of initial-state and
final-state radiation of photons was also devised.  These achievements
allowed a reduction of the uncertainty on $\sigma_{\rm WW}^{\rm theo}$ to
the current level of 0.5\%~\cite{Grunewald:2000ju,sigmaWtheo}.

The branching fractions of W bosons are derived from the number of
events measured in the different
channels~\cite{unknown:2003ih,sigmaW}. First, the branching fractions
into the three different lepton families are determined, without the
assumption of lepton universality, as:

\begin{center}
  \begin{tabular}{lcl}
    $\rm Br(W\rightarrow e \bar{\nu}_e) $ &=& $ 10.59 \pm 0.17 \%$ \\
    $\rm Br(W\rightarrow \mu \bar{\nu}_\mu) $ &=& $ 10.55 \pm 0.16 \%$\\
    $\rm Br(W\rightarrow \tau \bar{\nu}_\tau) $ &=& $ 11.20 \pm 0.22 \%.$\\
  \end{tabular}
\end{center}

\noindent The three values are compatible, and assuming lepton
universality the branching fraction into hadrons is derived as:
\begin{displaymath}
\rm Br(W\rightarrow q \bar{q}') = 67.77\pm 0.28 \%.
\end{displaymath}
These results are in agreement with the Standard Model
predictions. The branching fraction of W bosons into hadrons depends
on the six elements $|V_{\rm qq'}|$ of the Cabibbo-Kobayashi-Maskawa
matrix not involving top quarks. LEP measurements provide an estimate
of the less-known $|V_{\rm cs}|$ element as~\cite{unknown:2003ih}:
\begin{displaymath}
|V_{\rm cs}|= 0.989 \pm 0.014.
\end{displaymath}
%

%%%%%%%%%%%%%%%%%%%%%%%%%%%%%%%%%%%%%%%%%%%%%%%%%%%%%%%%%%%%%%%%%%
\section*{Single W-boson production}
%%%%%%%%%%%%%%%%%%%%%%%%%%%%%%%%%%%%%%%%%%%%%%%%%%%%%%%%%%%%%%%%%%

\begin{figure}
  \begin{center}
    \includegraphics[width=\textwidth]{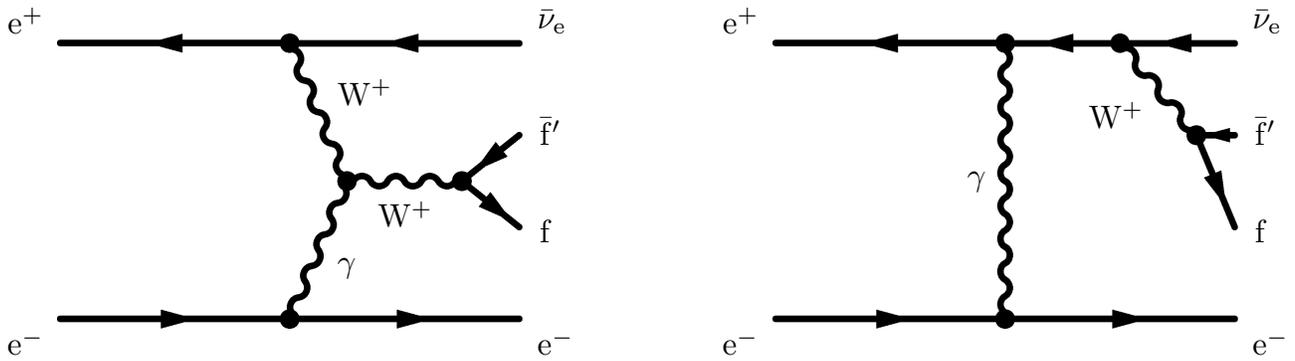}
    \caption{\label{fig:W2} Some of the Feynman diagrams describing
    single W-boson production at LEP.}
  \end{center}
\end{figure}

\begin{figure}
  \begin{center}
    \includegraphics[width=\textwidth]{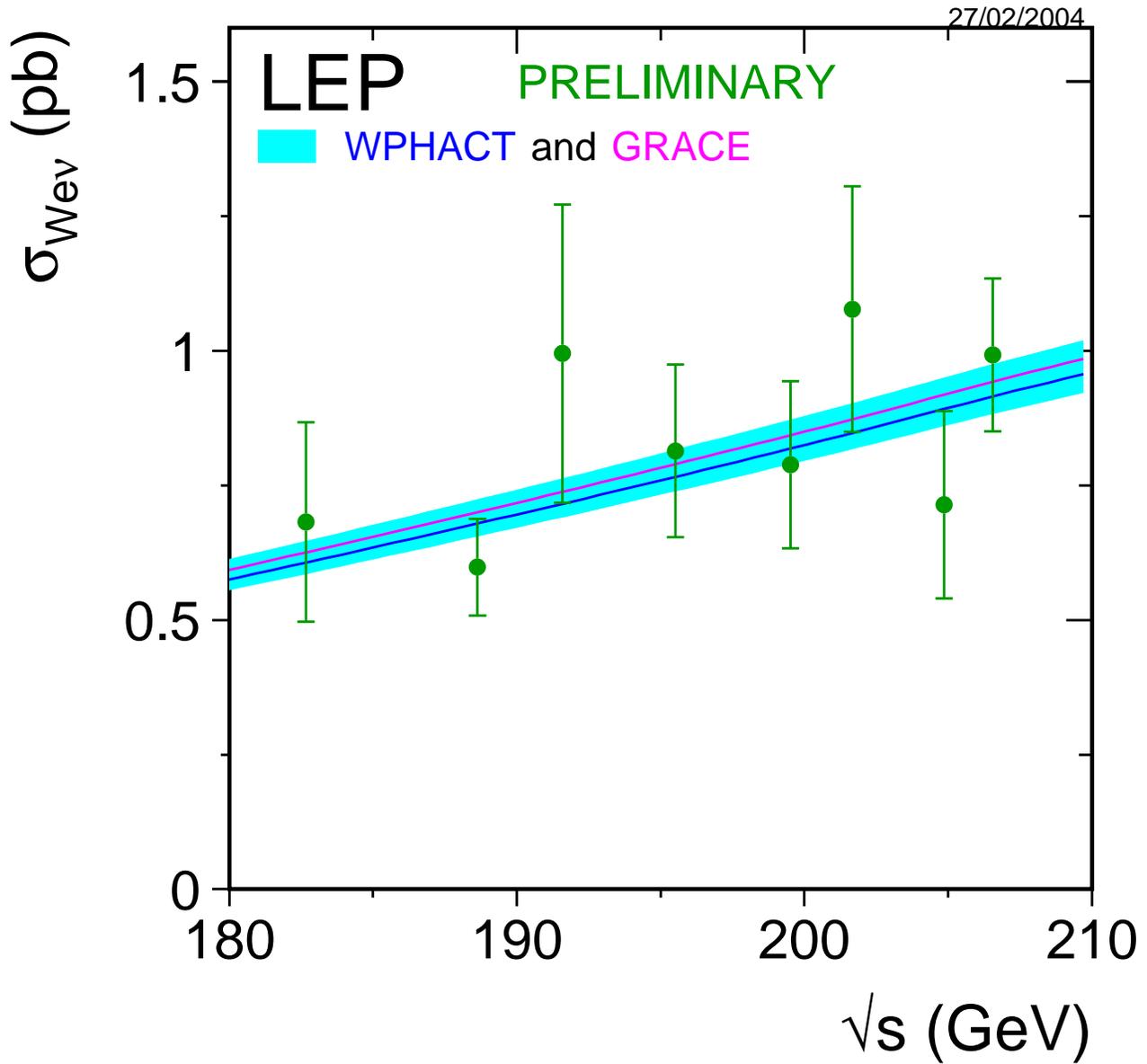}
    \caption{\label{fig:W4}  Measurement of the cross section for
    single W-boson production at LEP as a function of
    $\sqrt{s}$. Standard Model predictions~\cite{singleWtheo} are indicated by the band,
    whose width represent the theoretical uncertainty of 5\%~\cite{Grunewald:2000ju}.}
  \end{center}
\end{figure}

The gradual increase of the LEP centre-of-mass energy from 130~GeV up
to 209~GeV provided unique conditions to search for manifestations of
New Physics beyond the Standard Model. The production of particles
predicted by Supersymmetry, for example, would result in striking
signatures, such as events with two hadronic jets and large missing
energy, due to the production of weakly-interacting, and hence
undetected, neutralinos. Surprisingly, such events were found in the
LEP data. However, with an invariant mass of the hadronic system close
to $m_{\rm W} $ they were ascribed to hadronic decays of W bosons
single-produced through the process $\rm e^+e^- \rightarrow W^+ e^-
\bar{\nu}_e $, rather than to a discovery of Supersymmetry. This
process is described by Feynman diagrams like those presented in
Figure~\ref{fig:W2}, where the electrons escape detection as they are
scattered inside, or close to, the beam pipe. The other signature of
single W-boson production is a single charged-lepton in an otherwise
empty event. After the first observation of this
process~\cite{Acciarri:1997wf}, around 700 events were selected by the
four LEP collaborations~\cite{singleW,Achard:2002vd}.
Figure~\ref{fig:W4} presents the results of a combination of the
measured cross sections~\cite{unknown:2003ih}. A good agreement with
the Standard Model predictions~\cite{singleWtheo} is observed, as
quantified by the ratio:
\begin{displaymath}
{\cal R}^{\rm We\nu}={\sigma_{\rm We\nu}^{\rm meas} \over \sigma_{\rm
We\nu}^{\rm theo}}=0.978\pm 0.080,
\end{displaymath}
where the uncertainty is mainly statistical. The calculation of
$\sigma_{\rm We\nu}^{\rm theo}$ is made difficult by the low-angle
scattering of the final-state electron and is assigned an uncertainty
of 5\%~\cite{Grunewald:2000ju}.

As shown in Figure~\ref{fig:W2}, single W-boson production is
sensitive to the $\gamma$WW coupling  and hence to the
electromagnetic properties of W bosons. The W-boson
magnetic dipole moment, $\mu_{\rm W}$, and electric quadrupole moment,
$q_{\rm W}$, are written as~\cite{Wmoments}:
\begin{equation}
\label{eq:W1}
\mu_{\rm W}= {e \over 2 m_{\rm
    W}}(1+\kappa_\gamma+\lambda_\gamma)\,\,\,\,\,\,\,\,\,\, q_{\rm W} = - {e
    \over m^2 _{\rm W}}(\kappa_\gamma-\lambda_\gamma),
\end{equation}
where $e$ is the electron charge and the parameters $\kappa_\gamma$
and $\lambda_\gamma$ describe the coupling of photons and W bosons. In
the Standard Model, $\kappa_\gamma=1$ and $\lambda_\gamma=0$.
Higher-order contributions are well below the statistical precision of
LEP~\cite{yellow2} data. As an example, a fit to the measured cross
section of single W-boson production yields~\cite{Achard:2002vd}:
\begin{displaymath}
\kappa_\gamma=1.12 \pm 0.11,
\end{displaymath}
in agreement with the Standard Model predictions. The uncertainty is
in equal parts statistical and systematic, the latter being mostly due
to the control of signal modelling and instrumental
effects.

%%%%%%%%%%%%%%%%%%%%%%%%%%%%%%%%%%%%%%%%%%%%%%%%%%%%%%%%%%%%%%%%%%
\section*{W-boson polarisation}
%%%%%%%%%%%%%%%%%%%%%%%%%%%%%%%%%%%%%%%%%%%%%%%%%%%%%%%%%%%%%%%%%%

The spin of W bosons has a transverse and a longitudinal helicity
component. The measurement of W-boson polarisation is of particular
interest since the longitudinal helicity component arises from
the mechanism of electroweak symmetry breaking which gives the W boson
its non-vanishing mass. Moreover, a comparison of the helicity fractions of
the $\rm W^-$ and $\rm W^+$ bosons allows a test of CP conservation.

The fractions of the three helicity states of W bosons produced in
$\rm e^+e^-$ collisions are a function of both $\sqrt{s}$ and the
cosine of the $\rm W^-$ production angle with respect to the electron
beam, $\cos\theta_{\rm W^-}$. For the data sample under investigation,
Monte Carlo programs~\cite{Jadach:2001mp} predict a longitudinal
polarisation of 24\%.

The polarisation of pair-produced W bosons is probed by reconstructing
the direction in which their decay products are emitted. The
experimental analyses are restricted to semi-leptonic events, where
the charge of the lepton defines the charge of the W bosons. Denoting
the fraction of the helicity states $-1$, $+1$ and 0 of $\rm W^-$
bosons as $f_-$, $f_+$ and $f_0$, the rest-frame lepton angular
spectrum in leptonic $\rm W^-$ decays is given by\footnote{Assuming CP
invariance, $f_-$, $f_+$ and $f_0$ also represent the fractions of the
helicity states $+1$, $-1$ and 0 of $\rm W^+$ bosons, respectively.}:
\begin{equation}
\label{eq:W2}
{1 \over N} {{\rm d}N \over {\rm d}\cos\theta^*_\ell} = f_- {3 \over
8} \left( 1+\cos\theta^*_\ell\right)^2 +f_+ {3 \over 8} \left(
1-\cos\theta^*_\ell\right)^2 +f_0 {3 \over 4} \sin^2\theta^*_\ell.
\end{equation}
As the quark charge is difficult to reconstruct, the rest-frame angular spectrum in hadronic
decays is folded as: 
\begin{equation}
\label{eq:W3}
{1 \over N} {{\rm d}N \over {\rm d}|\cos\theta^*_{\rm q}|} = f_{\pm}
{3 \over 4} \left( 1+|\cos\theta^*_{\rm q}|\right)^2 +f_0 {3 \over 2}
\left( 1-|\cos\theta^*_{\rm q}|\right)^2,
\end{equation}
where $f_{\pm}=f_+ + f_-$.

\begin{figure}
  \begin{center}
    \begin{tabular}{cc}
      \mbox{\includegraphics[width=0.5\textwidth]{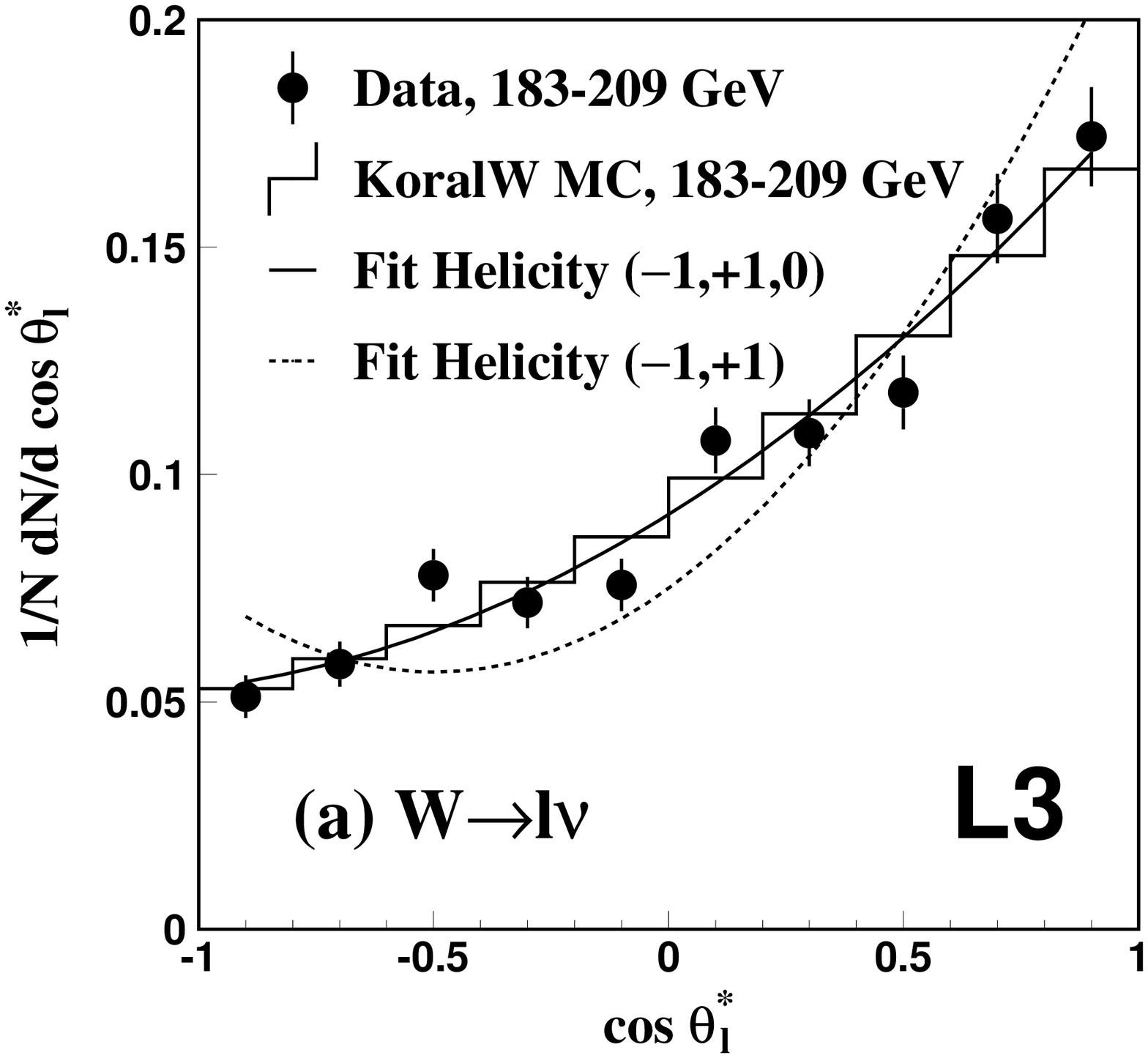}}&
      \mbox{\includegraphics[width=0.5\textwidth]{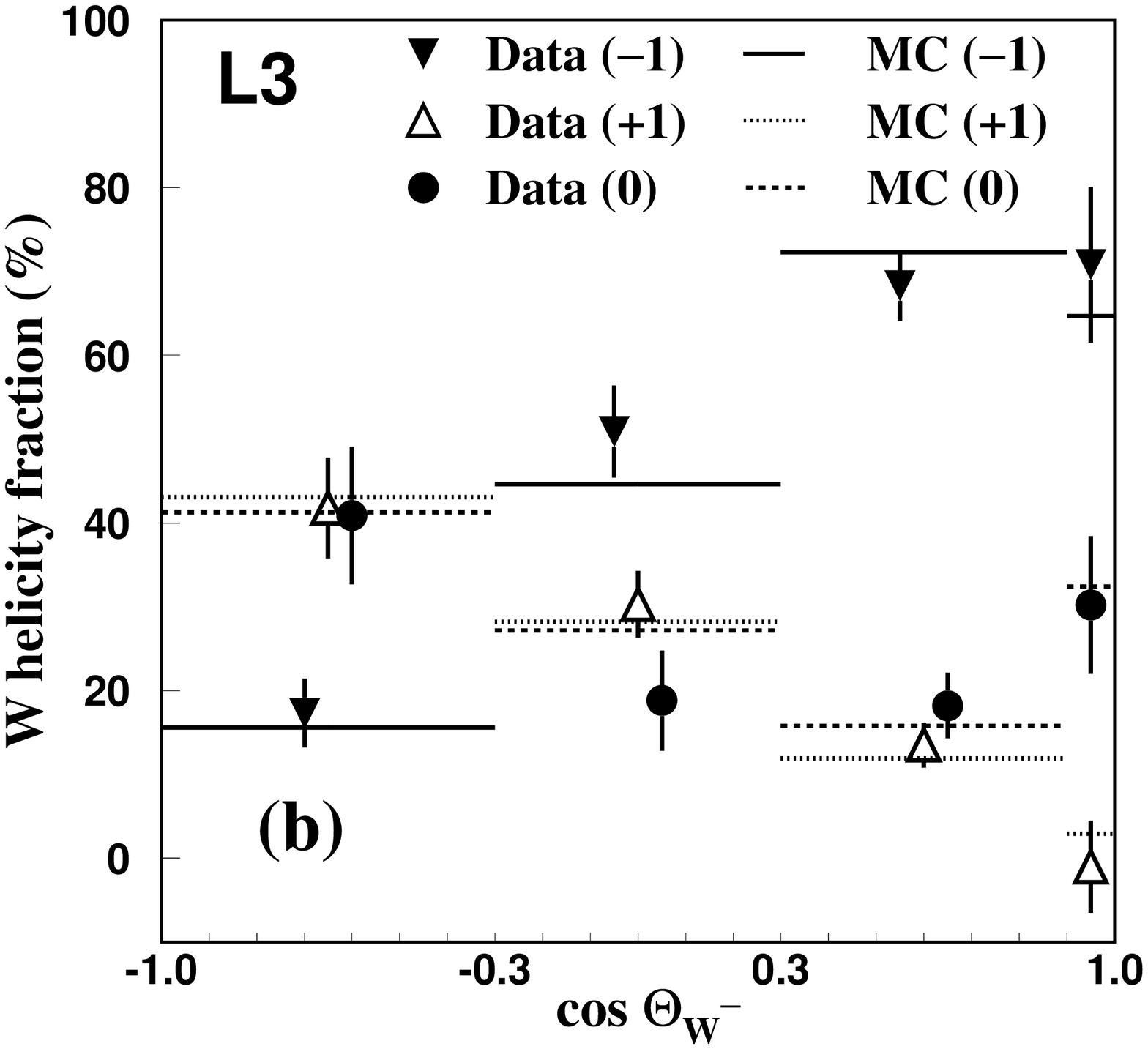}}\\
    \end{tabular}
    \caption{\label{fig:W5} a) Rest-frame lepton angular spectrum
    observed in data compared with Standard Model Monte Carlo
    expectations.  Results of fits with two and three helicity states
    are shown. A fit with no longitudinal polarisation fails to
    describe the data. b) Helicity fractions of W bosons measured as a
    function of the cosine of the W$^-$ polar angle.}
  \end{center}
\end{figure}

The L3 collaboration performed a fit of Equations~(\ref{eq:W2})
and~(\ref{eq:W3}) to about 2\,000 semi-leptonic
events~\cite{Achard:2002bv}. As shown in Figure~\ref{fig:W5}a, a fit
without  longitudinal polarisation fails to
describe the data. A fit with the three helicity components measures the fraction of
longitudinal polarisation to be in agreement with the predictions,
with a value:
\begin{displaymath}
f_0 = 21.8 \pm 3.1 \%,
\end{displaymath}
where the uncertainty is mainly statistical. The helicity fractions
are also measured in four different bins of $\cos\theta_{\rm W^-}$. A
good agreement with the predictions is found, as shown in
Figure~\ref{fig:W5}b. CP conservation is verified by separately
measuring the helicity fractions for W$^+$ and W$^-$ bosons, which are
found to be in agreement, with a statistical accuracy of about 30\%.

The OPAL collaboration determined the helicity fractions
through the investigation of the W-boson spin-density
matrix~\cite{Abbiendi:2003wv}. The elements of this matrix are defined
as~\cite{Gounaris:1992kp}:
\begin{displaymath}
\rho^{\rm W^-}_{\tau\tau'}(s,\cos\theta_{\rm W^-}) = {
\sum_{\lambda,\lambda'} F^{(\lambda,\lambda')}_\tau
\left(F^{(\lambda,\lambda')}_{\tau'}\right)^\star \over
\sum_{\lambda,\lambda',\tau} \left| F^{(\lambda,\lambda')}_\tau
\right|^2 },
\end{displaymath}
where $F^{(\lambda,\lambda')}_\tau$ is the helicity amplitude to
produce a W$^-$ boson with helicity $\tau$ from an electron with
helicity $\lambda$ and a positron with helicity $\lambda'$. The
spin-density matrix is a Hermitian matrix with  unit trace described by
eight free parameters. The $\rho_{++}$, $\rho_{--}$ and $\rho_{00}$
diagonal elements correspond to the fractions $f_+$, $f_-$ and $f_0$,
respectively. The $\rho_{\tau\tau'}$ elements are derived from the
measurement of the projection operators $\Lambda_{\tau\tau'}$. These
are known functions of the polar and azimuthal rest-frame angles of
the final-state fermions and project the differential cross section
for W-boson pair-production onto the $\rho_{\tau\tau'}$
elements~\cite{Bilenky:1993ms}. By studying a sample of about 4\,000
semi-leptonic events, the value:
\begin{displaymath}
f_0 = 23.9 \pm 2.4 \%
\end{displaymath}
is obtained, where the uncertainty is mainly statistical.  Compatible preliminary
results were also reported by the DELPHI
Collaboration~\cite{Delphi2003052}.

CP invariance implies $\rho^{\rm W^-}_{\tau\tau'}=\rho^{\rm W^
+}_{-\tau-\tau'}$~\cite{Gounaris:1991ce}. Introducing the pseudo
time-reversal operator $\hat {\rm T}$, which transforms the helicity
amplitudes into their complex conjugates and is equivalent to the T
operator at tree level~\cite{Hagiwara:1986vm}, CP$\hat {\rm T}$
invariance implies $\rho^{\rm W^-}_{\tau\tau'}=\left(\rho^{\rm W^
+}_{-\tau-\tau'}\right)^\star$. Therefore, at tree level, only the
imaginary parts of the $\rho_{\tau\tau'}$ elements are sensitive to
possible CP violation. By introducing the cross sections:
\begin{displaymath}
\sigma^{\rm W^\pm}_{\tau\tau'} = \int^{+1}_{-1} \Im\left\{ \rho^{\rm
W^\pm}_{\tau\tau'}\right\} {{\rm d}\sigma \over {\rm d}\cos\theta_{\rm
W^-}} {\rm d}\cos\theta_{\rm W^-},
\end{displaymath}
three quantities sensitive to tree-level CP violation are formed as:
\begin{displaymath}
\Delta^{\rm CP}_{+-} = \sigma^{\rm W^-}_{+-} -\sigma^{\rm
W^+}_{-+}\,\,\,\,\,\,\,\,\,\, \Delta^{\rm CP}_{+0} = \sigma^{\rm W^-}_{+0}
-\sigma^{\rm W^+}_{-0}\,\,\,\,\,\,\,\,\,\, \Delta^{\rm CP}_{-0} = \sigma^{\rm
W^-}_{-0} -\sigma^{\rm W^+}_{+0},
\end{displaymath}
as well as three quantities sensitive to loop effects:
\begin{displaymath}
\Delta^{\rm CP\hat T}_{+-} = \sigma^{\rm W^-}_{+-} +\sigma^{\rm
W^+}_{-+}\,\,\,\,\,\,\,\,\,\, \Delta^{\rm CP\hat T}_{+0} = \sigma^{\rm W^-}_{+0}
+\sigma^{\rm W^+}_{-0}\,\,\,\,\,\,\,\,\,\, \Delta^{\rm CP\hat T}_{-0} =
\sigma^{\rm W^-}_{-0} +\sigma^{\rm W^+}_{+0}.
\end{displaymath}
The measured values of all these quantities are compatible with zero
within a statistical accuracy of about 15\% and no effects
of CP violation are observed~\cite{Abbiendi:2003wv}.  Compatible
preliminary results were also reported by the L3
Collaboration~\cite{L32793}.

%%%%%%%%%%%%%%%%%%%%%%%%%%%%%%%%%%%%%%%%%%%%%%%%%%%%%%%%%%%%%%%%%%
\section*{Triple gauge-boson-couplings}
%%%%%%%%%%%%%%%%%%%%%%%%%%%%%%%%%%%%%%%%%%%%%%%%%%%%%%%%%%%%%%%%%%

The most general form for the $V$WW vertex, with $V$ denoting either a
photon or a Z boson, is  described by the effective
Lagrangian~\cite{Hagiwara:1986vm,Gaemers:1978hg}:
\begin{eqnarray}
\label{eq:W4}
i{\cal L}_{\rm eff}^{V{\rm WW}} & = & g_{V{\rm WW}}\, \Bigl[ g_1^V V^\mu \left(
W^{-}_{\mu\nu}W^{+\nu}-W^{+}_{\mu\nu}W^{-\nu}\right) \\ \nonumber
& &  +
\kappa_V\,  W^{+}_{\mu}W^{-}_{\nu}V^{\mu\nu} +{\lambda_V\over m_{\rm W}^2}\,
V^{\mu\nu}W^{+\rho}_{\!\!\nu}W^-_{\rho\mu} \\ \nonumber
& & +ig_5^V\varepsilon_{\mu\nu\rho\sigma}\Bigl(
(\partial^\rho W^{- \mu})W^{+\nu} -
W^{- \mu}(\partial^{\rho}W^{+\nu}) \Bigr) V^{\sigma} \\ \nonumber
& & +ig_4^V W^{+}_{\mu}W^-_{\nu} (\partial^\mu V^\nu+\partial^\nu V^\mu)\\ \nonumber
& & -\frac{\tilde \kappa_V}{2} W^{-}_{\mu}W^+_{\nu}
\varepsilon^{\mu\nu\rho\sigma}
V_{\rho\sigma} - 
{\tilde \lambda_V\over {2 m_{\rm W}^2}}\,
W^{-}_{\rho\mu}{W^{+\mu}}_{\nu}\varepsilon^{\nu\rho\alpha\beta}
V_{\alpha\beta}
\Bigr],
\end{eqnarray}
where $F_{\mu\nu}=\partial_\mu F_\nu - \partial_\nu F_\mu$.  Once the
overall couplings are defined as $g_{\gamma{\rm WW}} = e$ and $g_{\rm
ZWW} = e\cot\theta_w$, where $\theta_w$ is the weak mixing-angle,
seven complex parameters describe the ZWW vertex and seven more
describe the $\gamma$WW vertex. These are too many to be measured
simultaneously, and some hypotheses are introduced. First, the
CP-violating parameters $g_4^V$, $\tilde \kappa_V$ and $\tilde
\lambda_V$ are discarded, as supported by the tests of CP conservation
discussed above. In addition, electromagnetic gauge invariance is
assumed, fixing $g_1^\gamma=1$ and $g_5^{\rm Z} = 0$. The remaining
five couplings $g_1^{\rm Z}$, $\kappa_\gamma$, $\kappa_{\rm Z}$,
$\lambda_\gamma$, $\lambda_{\rm Z}$ are assumed to be real. Their
Standard Model tree-level values are $g_1^{\rm Z} = \kappa_\gamma =
\kappa_{\rm Z} = 1$ and $\lambda_\gamma = \lambda_{\rm Z} = 0$.

Custodial SU(2) symmetry~\cite{yellow2,Bilenky:1993ms,Gaemers:1978hg}
implies $\kappa_{\rm Z} = g_1^{\rm Z} - \tan\theta_w(\kappa_\gamma
-1)$ and $\lambda_{\rm Z}=\lambda_\gamma$ and reduces the
parametrisation of TGCs to three quantities: $g_1^{\rm Z}$,
$\kappa_\gamma$ and $\lambda_\gamma$. As presented in
Equation~(\ref{eq:W1}), $\kappa_\gamma$ and $\lambda_\gamma$ are related
to the W-boson electromagnetic properties~\cite{Wmoments}.

The differential cross section of W-boson pair-production exhibits a
strong dependence on $g_1^{\rm Z}$, $\kappa_\gamma$ and
$\lambda_\gamma$. For unpolarised initial states, summing over the
final-state fermion helicities, fixing $m_{\rm W}$ and neglecting
photon radiation, five angles describe  completely the phase space of
W-boson pair-production, $\Omega$, and are used for the TGC
determination. In addition to $\theta_{\rm W^-}$, these are the
rest-frame polar and azimuthal decay angles of the fermions from the
W$^-$ decays and of the anti-fermions from the W$^+$ decays.

For semi-leptonic events the determination of the charge of the W
bosons, crucial for the reconstruction of $\theta_{\rm W^-}$, is
accurate. These events also allow the identification of the fermion and
anti-fermion in the W-boson leptonic decay.  On the other side of the
event, no attempts to identify the fermion and the anti-fermion in
W-boson hadronic decays are usually made and folded angular
distributions are considered.

For fully-hadronic events, jet-charge techniques result in a
satisfactory tagging of the W-boson charge and allow to reconstruct
$\theta_{\rm W^-}$. Folded distributions are used for all rest-frame
decay angles.

\begin{figure}
  \begin{center}
    \includegraphics[width=\textwidth]{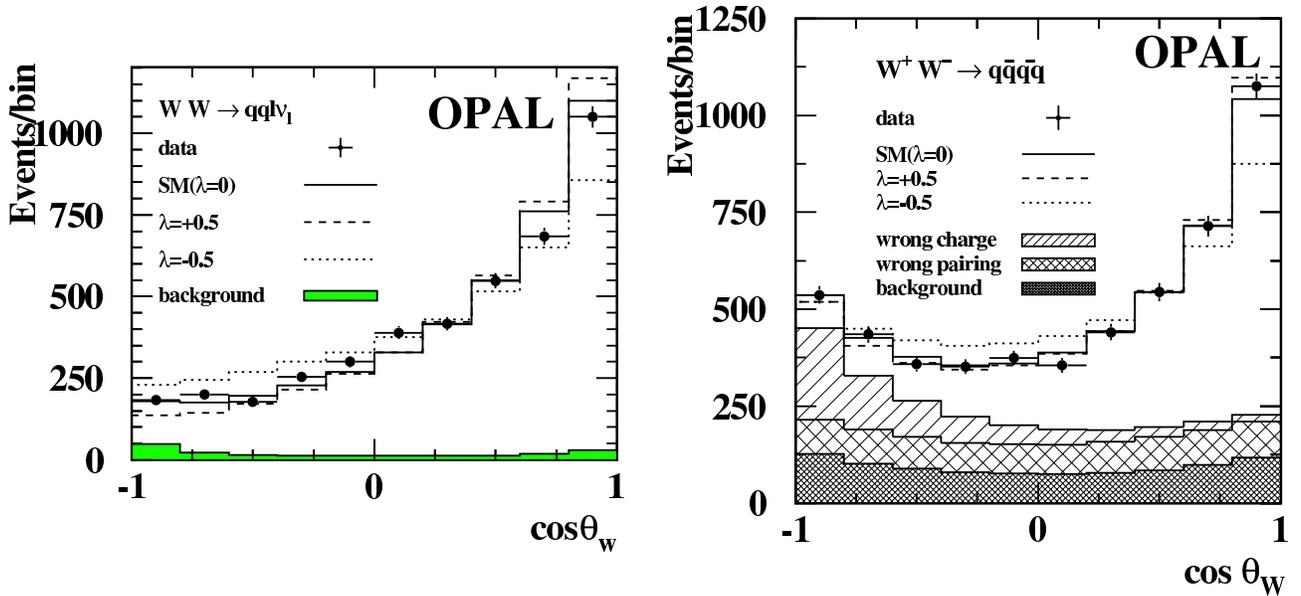}
    \caption{\label{fig:W6}Differential distribution for the cosine of
    the $\rm W^-$ polar angle for semi-leptonic and fully-hadronic
    events.  Predictions from the Standard Model and in presence of an
    anomalous value of the coupling $\lambda_\gamma$ are also given.}
  \end{center}
\end{figure}

The largest sensitivity to TGCs comes from $\cos\theta_{\rm
W^-}$. Figure~\ref{fig:W6} compares its distributions, as observed by
the OPAL collaboration, with the predictions for the Standard Model
value $\lambda_\gamma=0$ and for $\lambda_\gamma=\pm 0.5$.

\begin{figure}
  \begin{center}
    \begin{tabular}{cc}
      \mbox{\includegraphics[width=0.5\textwidth]{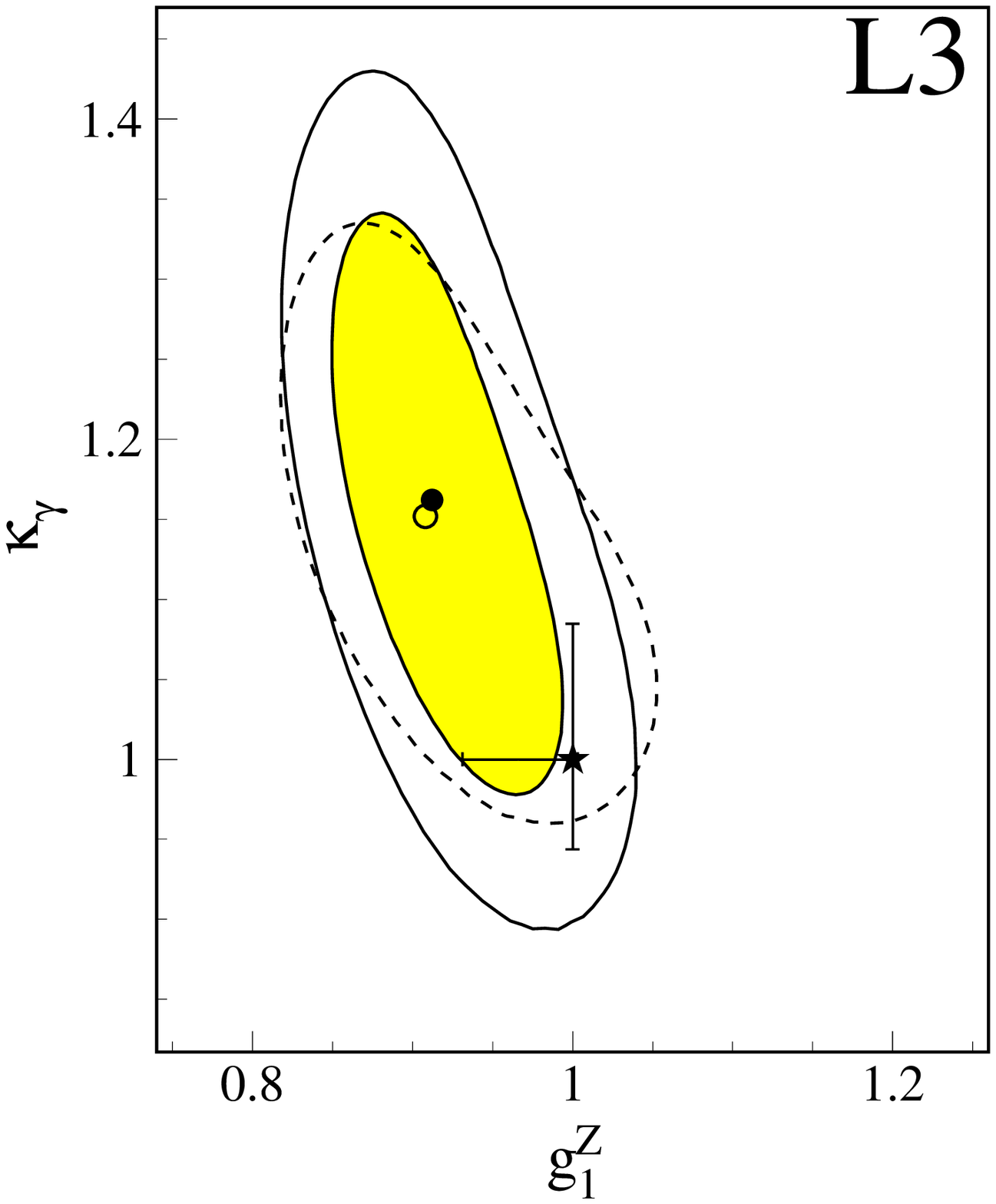}}&
      \mbox{\includegraphics[width=0.5\textwidth]{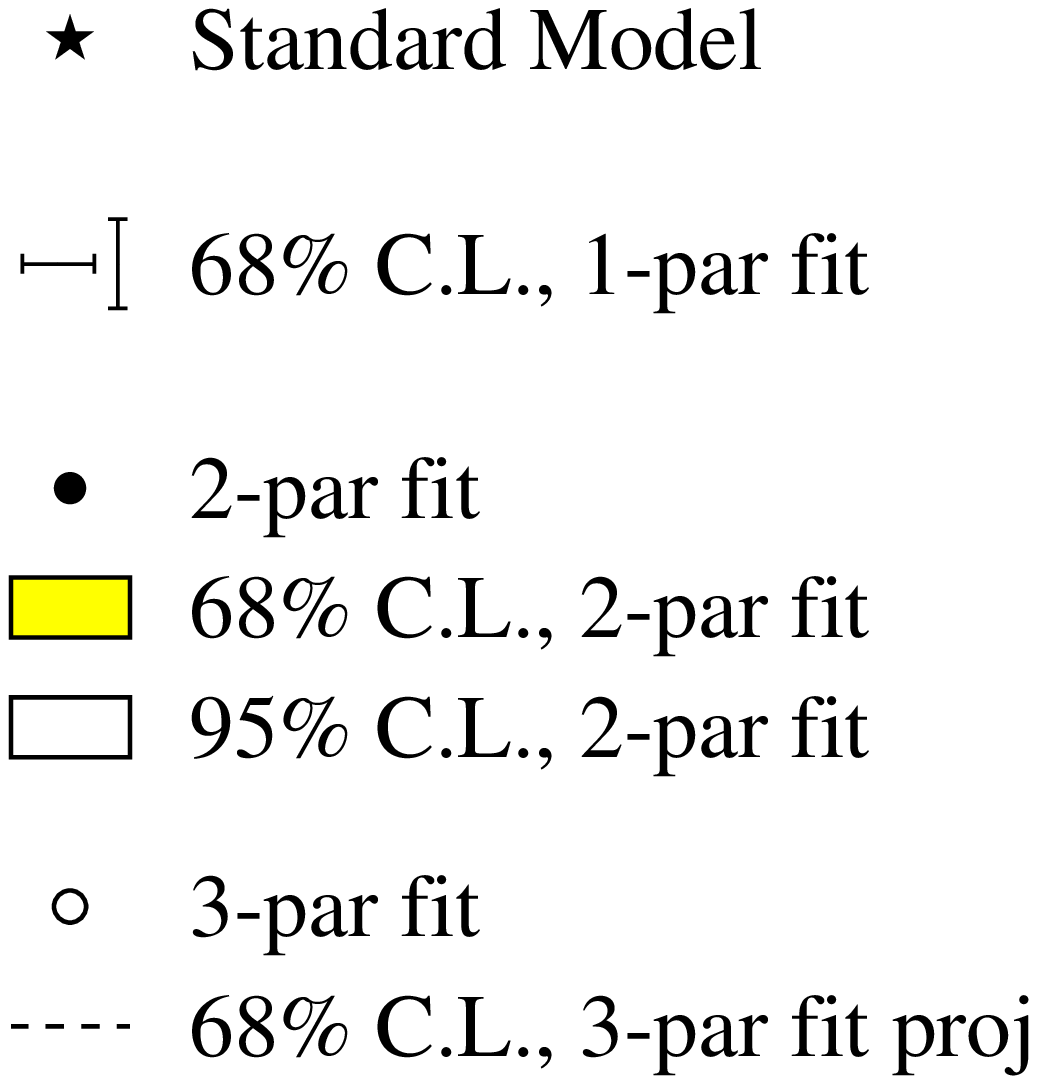}}\\
    \end{tabular}
    \caption{\label{fig:W7}Results of  one-, two- and
    three-dimensional determinations of the couplings $g_1^{\rm Z}$ and
    $\kappa_\gamma$.}
  \end{center}
\end{figure}

A method the determination of TGCs is to fit to the data the
five-dimensional differential cross section obtained by re-weighting
Monte Carlo events as a function of $g_1^{\rm Z}$, $\kappa_\gamma$ and
$\lambda_\gamma$~\cite{Achard:2004ji}.  Fits to each of the tree
couplings are performed as well as simultaneous fits to two or three
couplings.  As an example, Figure~\ref{fig:W7} presents the results of
the fits for $g_1^{\rm Z}$ and $\kappa_\gamma$.

TGCs are also determined with an ``optimal observable
analysis''~\cite{Diehl:1993br}. As the Lagrangian of
Equation~(\ref{eq:W4}) is linear in the TGCs, $\alpha_i$, the
differential cross section for W-boson pair-production is a
second-order polynomial function:
\begin{displaymath}
{{\rm d}\sigma(\Omega,\alpha_i) \over {\rm d}\Omega} =
S^{(0)}_{\phantom{i}}(\Omega)+\sum_i \alpha_i S^{(1)}_i(\Omega)
+\sum_{i,j} \alpha_i\alpha_j S^{(2)}_{ij}(\Omega),
\end{displaymath}
where  the functions $S^{(0)}_{\phantom{i}}$, $S^{(1)}_i$ and $S^{(2)}_{ij}$ are
known. All the information on the TGCs  is then summarised by the observables:
\begin{center}
  \begin{tabular}{rcl}
    ${\cal O}^{(1)}_i   $&=&$  S^{(1)}_i(\Omega)/S^{(0)}_{\phantom{i}}(\Omega)$ \\
    ${\cal O}^{(2)}_i   $&=&$  S^{(2)}_{ii}(\Omega)/S^{(0)}_{\phantom{i}}(\Omega)$\\
    ${\cal O}^{(2)}_{ij} = {\cal O}^{(2)}_{ji} $&=&$   S^{(2)}_{ij}(\Omega)/S^{(0)}_{\phantom{i}}(\Omega)$,\\
\end{tabular}
\end{center}
which are reconstructed from data and fit to determine the
TGCs~\cite{TGC}. 

Compatible results are found by all LEP collaborations and their
preliminary
combination gives~\cite{unknown:2003ih}:
\begin{displaymath}
g_1^{\rm Z} = 0.991^{+0.022}_{-0.021}\,\,\,\,\, 
\kappa_\gamma = 0.984^{+0.042}_{-0.047}\,\,\,\,\,
\lambda_\gamma = -0.016^{+0.021}_{-0.023}, 
\end{displaymath}
in agreement with the Standard Model prediction. The uncertainties are
in equal part statistical and systematic. The latter follows from the
theoretical uncertainties on the description of the differential cross
sections of W-boson pair-production~\cite{Grunewald:2000ju}. Results
from two- and three-dimensional fits are also in agreement with the
Standard Model predictions.

These results also include information from a partial reconstruction
of fully-leptonic events and from single W-boson production and
single-photon production. The last phenomenon is mostly due to the
radiation of a photon in the initial state of the process $\rm
e^+e^-\rightarrow \nu\bar{\nu}$. However, it also receives a small
contribution from the $\rm e^+e^-\rightarrow \nu_e\bar{\nu}_e\gamma$
process where the photon is produced through W-boson fusion in a
$\gamma$WW vertex.  Semi-leptonic and fully-hadronic events from
W-boson pair-production are largely more sensitive than these other
processes.  In particular, they are two, ten and five times more
sensitive than fully-leptonic events for $g_1^{\rm Z}$,
$\kappa_\gamma$ and $\lambda_\gamma$, respectively, around ten times
more sensitive than single W-boson production for $\lambda_\gamma$,
and four and twenty times more sensitive than single-photon production
for $\kappa_\gamma$ and $\lambda_\gamma$, respectively. The only
exception is the comparable sensitivity of single W-boson production
to $\kappa_\gamma$.

If the W boson were an extended object, such as an ellipsoid of
rotation, its average radius $R_{\rm W}$ would be related to its
magnetic dipole moment and hence to the TGC as: $R_{\rm W} =
(\kappa_\gamma+\lambda_\gamma-1)/m_{\rm W}$~\cite{Brodsky:1980zm}.
The measurements~\cite{Achard:2004ji} indicate that W bosons
are point-like particles down to a scale of $10^{-19}$~m:
\begin{displaymath}
     R_{\rm W} = (0.3\pm1.9)\times 10^{-19}~\mathrm{m}.
\end{displaymath}

%%%%%%%%%%%%%%%%%%%%%%%%%%%%%%%%%%%%%%%%%%%%%%%%%%%%%%%%%%%%%%%%%%
\section*{W-boson mass}
%%%%%%%%%%%%%%%%%%%%%%%%%%%%%%%%%%%%%%%%%%%%%%%%%%%%%%%%%%%%%%%%%%

Early measurements of $m_{\rm W}$ at LEP were performed with about
10~pb$^{-1}$ of data collected by each experiment at the W-boson
pair-production threshold, where the cross section depends strongly on $m_{\rm
W}$~\cite{threshold}.
The combined result reads~\cite{unknown:2003ih}:
\begin{displaymath}
m_{\rm W}^{\rm threshold} = 80.40 \pm 0.21\, {\rm GeV},
\end{displaymath}
where the uncertainty is mainly statistical.

\begin{figure}
  \begin{center}
    \begin{tabular}{cc}
      \mbox{\includegraphics[width=0.5\textwidth]{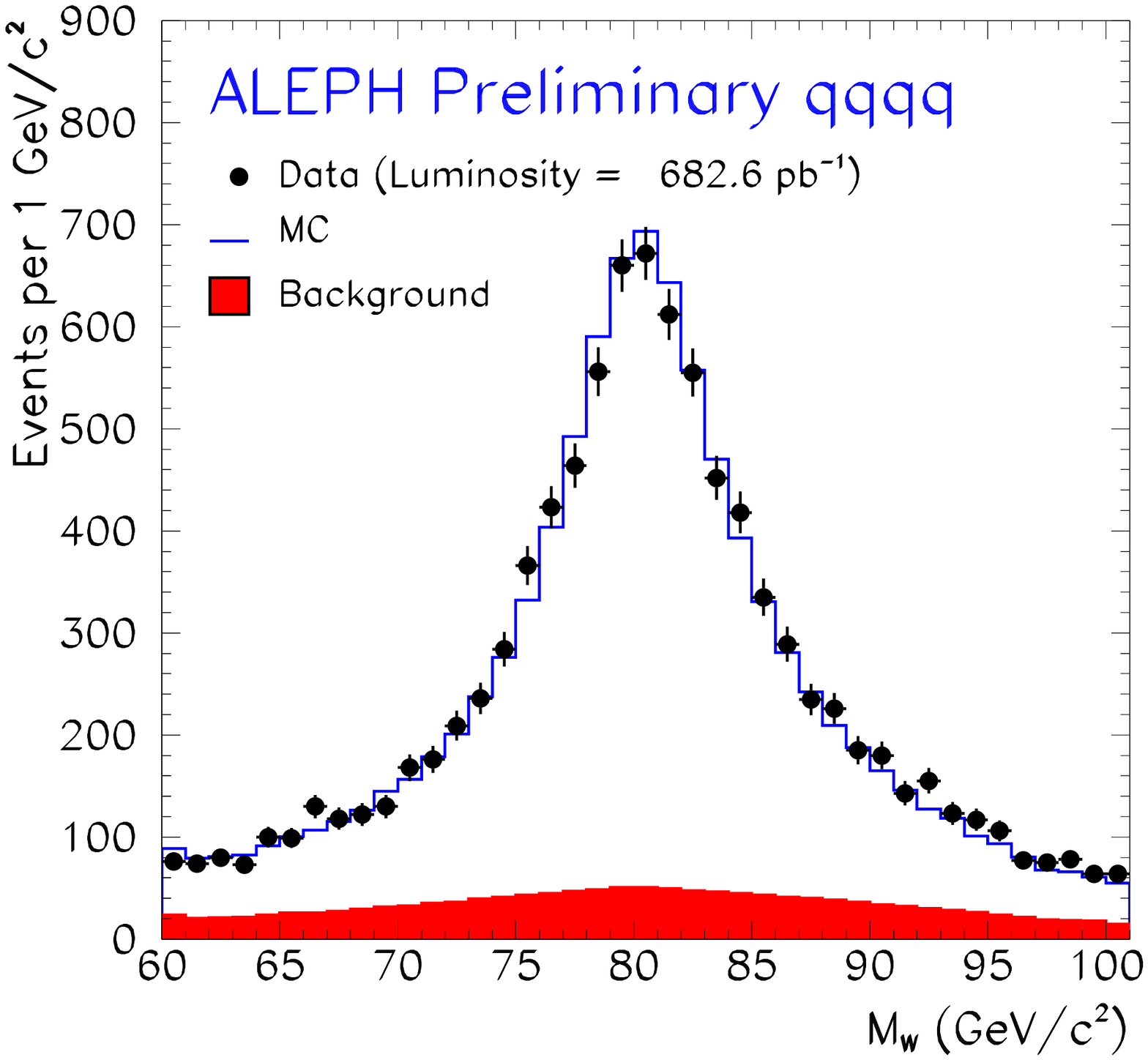}}&
      \mbox{\includegraphics[width=0.5\textwidth]{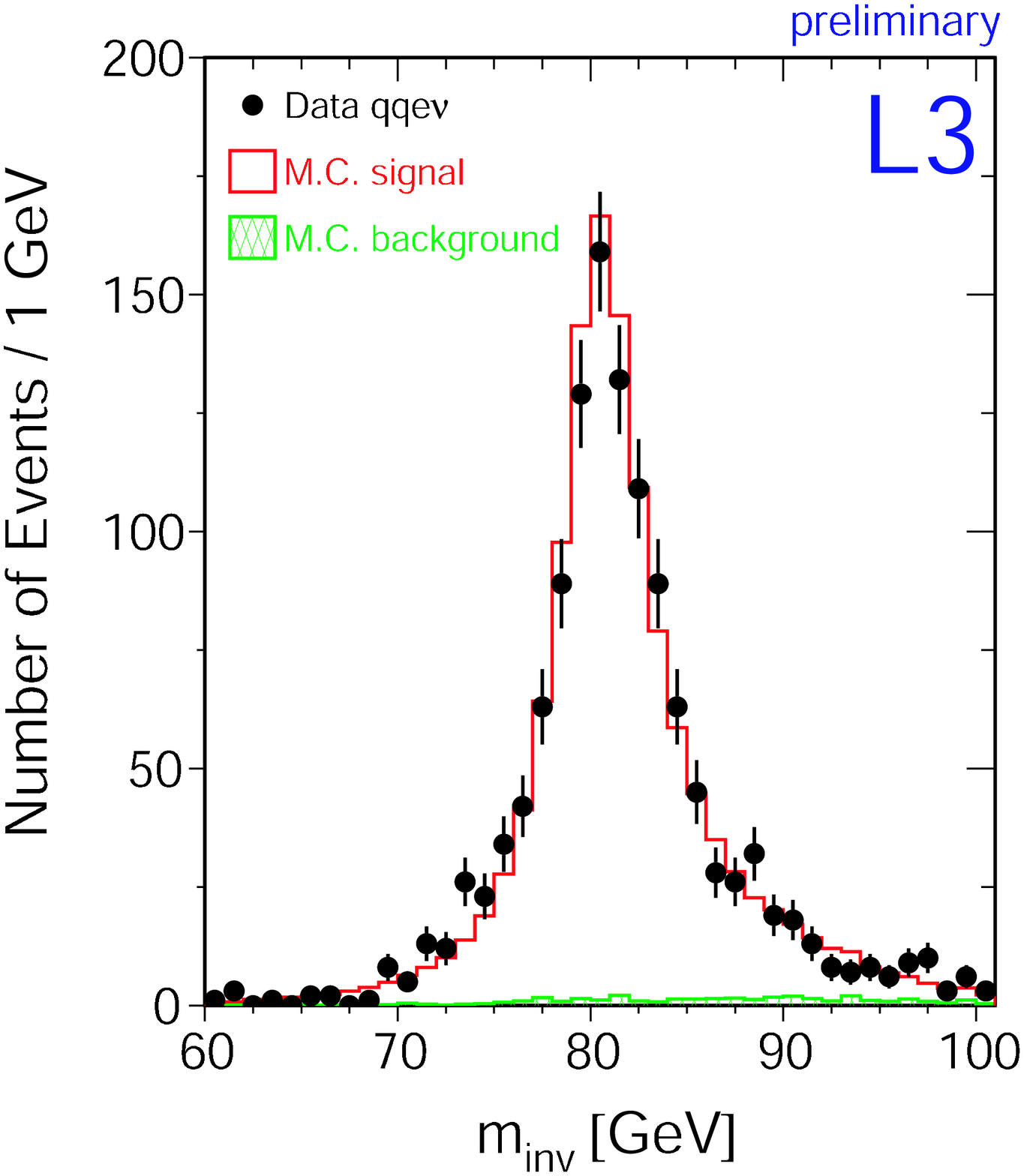}}\\
      \mbox{\includegraphics[width=0.5\textwidth]{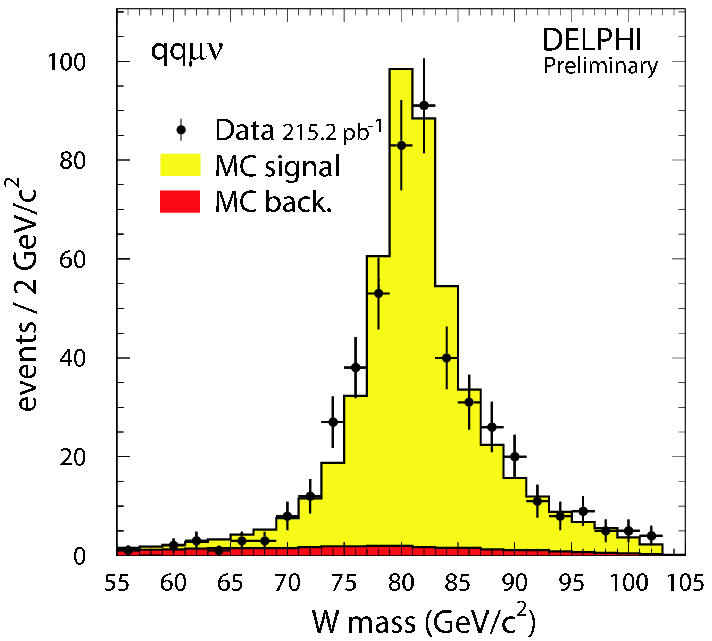}}&
      \mbox{\includegraphics[width=0.5\textwidth]{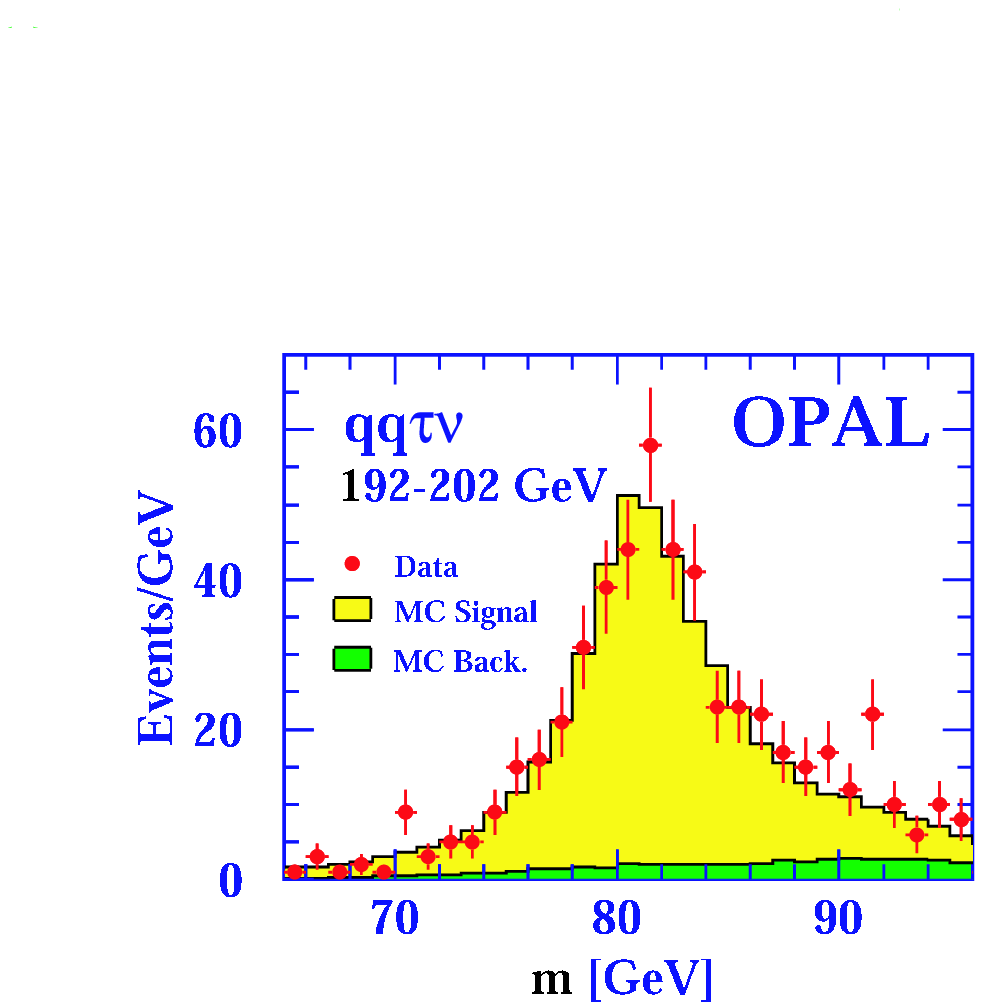}}\\
    \end{tabular}
    \caption{\label{fig:W8}W-boson mass spectra reconstructed for
    fully-hadronic events and semi-leptonic events with electrons,
    muons and tau leptons.}
  \end{center}
\end{figure}

Higher LEP centre-of-mass energies allow direct reconstruction of W
bosons and mass spectra are obtained for fully-hadronic and
semi-leptonic events, as shown in Figure~\ref{fig:W8}. Three
techniques were developed to extract $m_{\rm W}$ from these spectra or
from related quantities. A first technique is to fit the reconstructed
$m_{\rm W}$ spectrum with a Breit-Wigner function, using detector
resolutions obtained from Monte Carlo simulations. A second technique
is to compare directly Monte Carlo simulations which include all known
detector effects and physical processes to the data. Re-weighting
techniques are used to obtain simulations which are a function of
$m_{\rm W}$. A fit indicates the value of $m_{\rm W}$ which best describes
the data.  A last technique convolves all known detector and
physical processes to obtain a probability-density function for $m_{\rm
W}$ and a likelihood analysis of the data indicates the most probable value
of $m_{\rm W}$. A combination~\cite{unknown:2003ih} of the preliminary
results of the four experiments~\cite{mW} yields:
\begin{displaymath}
m_{\rm W}^{\rm qq\ell\nu} = 80.411 \pm 0.032\, ({\rm stat.}) \pm 0.030\,
({\rm syst.})\, {\rm GeV}
\end{displaymath}
\begin{displaymath}
m_{\rm W}^{\rm qqqq} = 80.420 \pm 0.035\,
({\rm stat.}) \pm 0.101\, ({\rm syst.})\, {\rm GeV}.
\end{displaymath}
for semi-leptonic and fully-hadronic final states, respectively, with
a correlation coefficient of 0.18. Their combination is:
\begin{displaymath}
m_{\rm W} = 80.412 \pm 0.029\, ({\rm stat.}) \pm 0.031\,
({\rm syst.})\, {\rm GeV}.
\end{displaymath}
This value also includes the results from the threshold measurements
and information from partial reconstruction of fully-leptonic
events~\cite{Abbiendi:2002ay}, which have a large statistical
uncertainty and hence little impact on the combined value. Analysis
methods similar to those used for the determination of $m_{\rm W}$ are
also used for the determination of the W-boson width, $\Gamma_{\rm
W}$, with the result:
\begin{displaymath}
\Gamma_{\rm W} = 2.150 \pm 0.091\, {\rm GeV}.
\end{displaymath}

  \begin{table}[h]
\begin{center}
    \begin{tabular}{|l|c|c|c|}
      \cline{2-4} \multicolumn{1}{c|}{} &
      \multicolumn{3}{c|}{Uncertainties on $m_{\rm W}$ (MeV)} \\ \hline
      Source of systematics &  Semi-leptonic & Fully-hadronic & Combined \\
      \hline 
      Bose-Einstein correlations & $-$& \phantom{0}35 & \phantom{0}3 \\
      Colour reconnection & $-$& \phantom{0}90 & \phantom{0}9 \\
      Beam energy & 17 & \phantom{0}17 & 17 \\
      Hadronisation & 19 & \phantom{0}18 & 18 \\ 
      Detector & 14 &      \phantom{0}10 & 14 \\ 
      Other & \phantom{0}8 & \phantom{00}9 & \phantom{0}8 \\ 
      \hline
      Total systematic & 31 & 101 & 31 \\ \hline 
    \end{tabular}
    \caption{\label{tab:W1}Sources of systematic uncertainty in the
      determination of $m_{\rm W}$ for hadronic and semi-leptonic final
      states and their combination.}
\end{center}
  \end{table}

The effects of different sources of systematic uncertainty are listed
in Table~\ref{tab:W1}~\cite{unknown:2003ih} and discussed in the
following.  Fully-hadronic events are affected by large systematic
uncertainties which are correlated among experiments and are due to
Bose-Einstein correlations (BEC) and colour reconnection (CR).  Their
weight in the combination is therefore only 10\%, sizably reducing the
statistical power of the analysis. At the time of writing, a
challenging program to reduce these uncertainties is in progress.

BEC are responsible for the enhancement of the production of pairs of
identical bosons close together in phase space. They were observed in
Z-boson decays and understood from quantum-mechanical principles. BEC
in W-boson decays were also observed and found to be similar to those
of Z-boson decays into light quarks~\cite{BEC}. The presence of BEC
between particles originating from decays of different W bosons would
modify the kinematics of the final-state particles and affect the
correspondence between $m_{\rm W}$ and the measured jet
four-momenta~\cite{fsi1,fsi2}. A large value for these ``inter-W''
BEC~\cite{Lonnblad:1997kk} would induce a shift on $m_{\rm W}$ of 35
MeV~\cite{unknown:2003ih}, assumed as systematic uncertainty in
Table~\ref{tab:W1}. In order to reduce this uncertainty, the LEP
collaborations have directly measured the amount of inter-W BEC by
comparing particle-correlation functions measured in fully-hadronic
events with those of four-jet events with no correlation. These are
obtained by superimposing the hadronic parts of two different
semi-leptonic events~\cite{BEC}. A combination of the results suggests
that only $(23\pm13)\%$ of the possible large effect is observed in
data~\cite{unknown:2003ih}. This result is compatible with little or
no BEC and reduces the present uncertainty on $m_{\rm
W}$ of 35~MeV to about 13~MeV.

The hadronisation of quarks from W-boson decays happens on a scale of
0.1~fm. Hadronic interactions, on the other hand, have a larger
characteristic distance of about 1~fm, which means that a substantial
cross-talk is possible between hadrons originating from different W
bosons. This process, called ``colour reconnection'', could modify the
four-momenta of the observed jets and introduce a shift in the
measured value of $m_{\rm W}$~\cite{fsi2,fsi3}. The LEP collaborations
have performed direct measurements of the extent of CR in
fully-hadronic events by comparing particle densities in regions
between jets originating from the same W bosons with those in the
regions between jets originating from different W bosons. Extreme CR
models are excluded~\cite{unknown:2003ih,CR}. This finding is
corroborated by the compatibility of the $m_{\rm W}$ determinations in
the semi-leptonic and fully-hadronic channels:
\begin{displaymath}
\Delta m_{\rm W} = m_{\rm W}^{\rm qq\ell\nu} - m_{\rm W}^{\rm qqqq} = -22 \pm 43\, {\rm MeV}.
\end{displaymath}
This data-driven estimate of CR effects is robust, but results into a
large range of possible shifts on $m_{\rm W}$, reflected by the 90~MeV
systematic uncertainty of Table~\ref{tab:W1}. A viable approach to
reduce this uncertainty is to make the measurement less sensitive to
CR effects. As CR mainly affects inter-jet regions, the LEP
collaborations are now modifying their clustering algorithms in order
to consider increasingly-narrow hadronic jets for the determination of
$m_{\rm W}$. Only events affected by a small uncertainty on CR will be
retained. Preliminary results indicate that this systematic
uncertainty can be halved at the price of an increase of 10\% of the
statistical uncertainty.

The determination of the LEP centre-of-mass energy is a source of
a systematic uncertainty which is correlated among all experiments and
channels. This follows from the use of $\sqrt{s}$ in kinematic fits
for the event reconstruction, resulting into an uncertainty on $m_{\rm
W}$ given by $\Delta m_{\rm W}/m_{\rm W} = \Delta
\sqrt{s}/\sqrt{s}$. Improved calibrations of the LEP centre-of-mass
energy~\cite{lepecal} will contribute to the reduction of this
uncertainty. As a cross check, the LEP experiments have reconstructed
the mass of the Z boson, $m_{\rm Z}$, using $\rm e^+e^- \rightarrow
Z\gamma$ events~\cite{Zgamma}. The results depend on $\sqrt{s}$. As
they are in agreement with the precision measurement of $m_{\rm Z}$
obtained from scans of the Z resonance, they validate the results of
the LEP energy calibration.

Another source of systematic uncertainty which is correlated among
experiment and channels, is the modelling of the hadronisation
process. As in the case of BEC and CR, possible changes in the
phase-space of the hadrons affect the $m_{\rm W}$
reconstruction. Different hadronisation models exist. They were
carefully ``tuned'' to reproduce experimental distributions observed
at LEP and at lower $\sqrt{s}$ and included in the Monte Carlo
simulations used to measure $m_{\rm W}$. A comparison of the results
obtained by using different hadronisation models results in an
uncertainty on $m_{\rm W}$ of 18~MeV.

The high precision of the $m_{\rm W}$ measurement calls for a detailed
understanding of detector resolutions and response. Uncertainties on
the energy scales of the calorimeters, and the angular determination
of leptons and jets directly affect $m_{\rm W}$. These uncertainties
may be sizable, but are not correlated among the experiments and are
largely diluted in the combination.

In conclusion, it is expected that the improved control of the
systematic uncertainties discussed above will increase the impact of
fully-hadronic events in the measurement of $m_{\rm W}$ and will thus
reduce the total uncertainty on $m_{\rm W}$ to around 35~MeV.

%%%%%%%%%%%%%%%%%%%%%%%%%%%%%%%%%%%%%%%%%%%%%%%%%%%%%%%%%%%%%%%%%%
\section*{Summary}
%%%%%%%%%%%%%%%%%%%%%%%%%%%%%%%%%%%%%%%%%%%%%%%%%%%%%%%%%%%%%%%%%%

The study of W-boson physics at LEP has been a success.  The existence
of the triple gauge-boson coupling ZWW was established, confirming
the non-Abelian structure of the Standard Model of the electroweak
interactions. W-boson longitudinal polarisation, which is a
consequence of the electroweak symmetry-breaking mechanism, was
observed and measured to be in agreement with the Standard Model
predictions. Within the available statistical precision, no hints of
CP violation in the W-boson system were found.

Several quantities describing fundamental properties of W bosons were
measured with an accuracy of a few percent: branching ratios, magnetic
dipole moment, electric quadrupole moment and couplings to Z bosons.
The mass and the width of the W bosons were measured to be:
\begin{displaymath}
m_{\rm W} = 80.412 \pm 0.042\, {\rm GeV}\,\,\,\,\,\,\,\,\,\,
\Gamma_{\rm W} = 2.150\pm 0.091\, {\rm GeV}. \\
\end{displaymath}
This value of $m_{\rm W}$ is in agreement with, and improves upon, the
measurements from hadron colliders. At the time of writing,
challenging studies aim to reduce the uncertainty on $m_{\rm W}$ to
around 35~MeV.

%%%%%%%%%%%%%%%%%%%%%%%%%%%%%%%%%%%%%%%%%%%%%%%%%%%%%%%%%%%%%%%%%%

%%%%%%%%%%%%%%%%%%%%%%%%%%%%%%%%%%%%%%%%%%%%%%%%%%%%%%%%%%%%%%%%%%

\newpage

\end{document}